\documentclass[journal=jacsat,manuscript=article]{achemso}

\usepackage{chemformula} 
\usepackage[T1]{fontenc} 
\usepackage{braket}
\usepackage[colorlinks=true,linkcolor=blue,citecolor=blue,urlcolor=blue]{hyperref}
\usepackage{etoc}

\author{Aurélie Champagne}
\affiliation[CNRS]{Institut de Chimie de la Matière Condensée de Bordeaux, CNRS, 33600 Pessac, France}
\alsoaffiliation[LBNL]{Materials Sciences Division, Lawrence Berkeley National Laboratory, Berkeley, CA 94720, USA}
\email{aurelie.champagne@icmcb.cnrs.fr}
\author{Olugbenga Adeniran}
\affiliation[Wayne]{Department of Chemistry, Wayne State University, Detroit, MI 48202, USA}
\author{Jonah B. Haber}
\affiliation[LBNL]{Department of Materials Science and Engineering, Stanford University, Stanford, CA 94305, USA}
\author{Antonios M. Alvertis}
\affiliation[NASA]{KBR, Inc., NASA Ames Research Center, Moffett Field, CA 94035, USA}
\alsoaffiliation[UTA]{Department of Physics, The University of Texas at Austin, Austin, TX 78712, USA}
\author{Zhen-Fei Liu}
\affiliation[Wayne]{Department of Chemistry, Wayne State University, Detroit, MI 48202, USA}
\author{Jeffrey B. Neaton}
\affiliation[LBNL]{Materials Sciences Division, Lawrence Berkeley National Laboratory, Berkeley, CA 94720, USA}
\alsoaffiliation[UCBerkeley]{Department of Physics, University of California, Berkeley, CA 94720, USA}
\alsoaffiliation[Kavli]{Kavli Energy Nanosciences Institute at Berkeley, Berkeley, CA 94720, USA}
\email{jbneaton@lbl.gov}

\title{Tunable electronic energy level alignment and exciton diversity in organic-inorganic van der Waals heterostructures}

\begin{document}

\begin{abstract}
van der Waals stacking of two-dimensional (2D) materials offers a powerful platform for engineering material interfaces with tailored electronic and optical properties. While most van der Waals multilayers have featured inorganic monolayers, incorporating molecular monolayers introduces new degrees of tunability and functionality. Here, we investigate hybrid bilayers composed of atomically thin perylene-based molecular crystals interfaced with monolayer transition metal dichalcogenides (TMDs), specifically MoS$_2$ and WS$_2$. Using \textit{ab initio} many-body perturbation theory within the GW approximation and the Bethe-Salpeter equation approach, we predict emergent properties beyond those of the isolated constituent systems. Notably, we find substantial renormalization of monolayer molecular crystal band gap due to TMD-induced polarization. Furthermore, by varying the TMD monolayer, we demonstrate tuning of the energy level alignment of the bilayer and subsequent control over a diversity of lowest-energy excitons, which include strongly bound hybrid excitons and long-lived charge-transfer excitons. These findings establish organic-inorganic van der Waals heterostructures as a promising class of materials for tunable optoelectronic devices and quantum excitonic phenomena, expanding the design space for low-dimensional systems.
\end{abstract}



\section{Introduction}

The unique electronic and optoelectronic properties of two-dimensional (2D) materials have boosted intensive interest in combining distinct 2D materials to create van der Waals (vdW) heterostructures for applications in optoelectronics, photovoltaics, and photodetection technologies~\cite{lo2011,novoselov2016,kafle2017,han2018,huang2018,kafle2019,liu032019,pei2020,xu2021}. Among these, heterostructures composed of monolayer transition metal dichalcogenides (TMDs) are particularly promising due to weak electronic screening, strong Coulomb interactions, and type-II energy level alignment (ELA), facilitating the formation of interlayer excitons (ILEs)~\cite{fogler2014,palummo2015, novoselov2016, latini2017,rivera2018,mak2018, ovesen2019,wilson2021,barre2022}. With a substantial binding energy and a much longer lifetime than intralayer excitons, ILEs can diffuse across longer distances, enabling facile exciton transport and coherent phenomena like exciton condensation~\cite{fogler2014,wang2019,gupta2020,liu2021}.

\begin{figure}[h!]
	\includegraphics[width=0.9\textwidth]{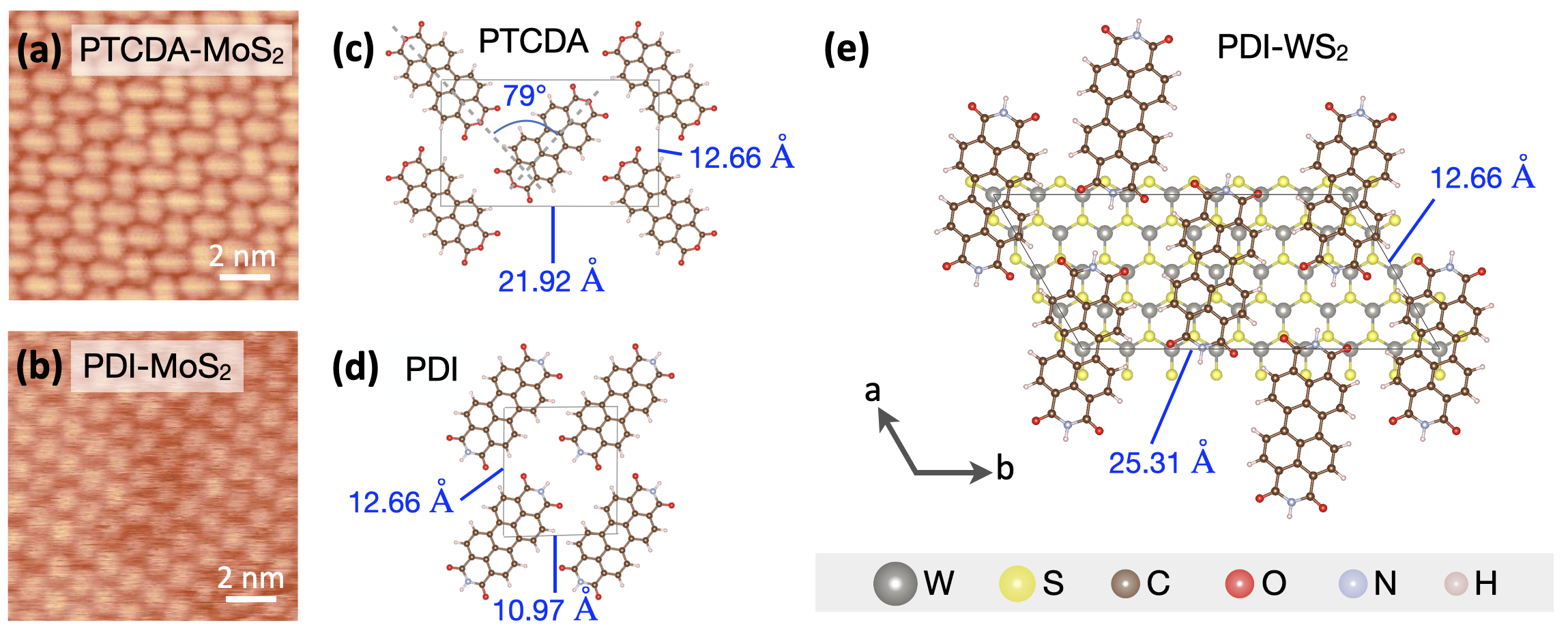}
	\centering
	\caption{Room-temperature STM image of (a) PTCDA and (b) PDI, grown on monolayer MoS$_2$ on a SiO$_2$ substrate, adapted from Ref.~\citenum{chowdhury2024}. Schematic atomic structure of (c) PTCDA $-$ with herringbone 2D arrangement,(d) PDI $-$ with brick-wall 2D arrangement, and (e) PDI-WS$_2$ interface supercell. Structural parameters are indicated with blue labels.}
	\label{fig1}
\end{figure}

The combinatorial space associated with vdW heterostructures is constrained by the variety of available 2D materials. Recently, organic molecules and their monolayers have emerged as compelling components in these systems due to their heightened sensitivity to their environment, which allows for tunable ELA and excitonic properties. Moreover, hybrid organic-inorganic heterostructures have potential applications in silicon solar cells with enhanced efficiency, by harnessing singlet exciton fission~\cite{einzinger2019,nagaya2024}. The spontaneous self-assembly of organic molecules on vdW surfaces further enhances the appeal of these heterostructures, and can enable highly ordered crystalline structures with preferred molecular orientations and stacking configurations that depend on the monomer~\cite{chowdhury2024}. For instance, planar perylene-tetracarboxylic dianhydride (PTCDA) molecules form ordered islands with a herringbone packing motif on monolayer TMDs~\cite{xu2021,zheng2016,chowdhury2024} (Figs.~\ref{fig1}a and \ref{fig1}c), while perylene diimide (PDI) molecules adopt a brick-wall arrangement~\cite{chowdhury2024} (Figs.~\ref{fig1}b and \ref{fig1}d). Recent studies have reported strong correlations between monomer composition and layer packing for a series of perylene derivatives, using both experimental and computational approaches~\cite{chowdhury2024,wang2025}.

Despite their promise, organic-inorganic vdW heterostructures remain relatively rare and largely underexplored. Only a few studies have demonstrated type-II interfaces in such systems, including phthalocyanine-TMD (ZnPc-MoS$_2$, H$_2$Pc-MoS$_2$, H$_2$Pc-WS$_2$)~\cite{adeniran2021,ulman2021}, and pentacene-black phosphorous hybrid interfaces~\cite{uddin2022}. In contrast to TMD bilayers $-$ where ILE formation is highly sensitive to the angular alignment of the bilayers~\cite{wang2019}, necessitating precise and complex fabrication techniques $-$ organic-inorganic heterostructures tend to exhibit momentum-direct, low-energy excitons due to the narrow bandwidth of molecular states~\cite{pei2020,ulman2021,xu2021}. Given that molecular systems typically host tightly bound Frenkel excitons while TMDs host more delocalized Wannier excitons, the nature of emergent excitons in such hybrid heterostructures is an active area of study. In fact, beyond conventional ILEs, these systems may host novel excitonic species such as hybrid or delocalized molecular excitons~\cite{krumland2021,oliva2022,bennecke2024,chowdhury2025}.

In this work, using state-of-the art \textit{ab initio} many-body perturbation theory~\cite{hedin1965,strinati1982,hybertsen1986,rohlfing2000} and the GW plus Bethe-Salpeter equation (GW-BSE) approach, we compute quasiparticle (QP) energies and optical excitations of four hybrid vdW bilayer systems comprising either PTCDA or PDI monolayers interfaced with MoS$_2$ or WS$_2$ monolayers (Fig.~\ref{fig1}e). We use an approximate, yet accurate, approach to effectively capture the dielectric screening in the heterogeneous environment while reducing the computational cost of the polarizability~\cite{liu2019}, a bottleneck in GW-BSE calculations for large systems. We combine efficient exciton decomposition~\cite{uddin2022} and visualization methods~\cite{sharifzadeh2013,sharifzadeh2015} to gain deep insight into the nature, localization, and charge-transfer character of the excited states that form in organic-TMD heterostructure bilayers (see SI for details). We find that TMD choice (MoS$_2$ vs. WS$_2$) changes the ELA from type-I to type-II, which in turn profoundly affects the nature of the lowest-energy excitons. Nonlocal polarization effects cause a dramatic renormalization of molecular monolayer band gap by up to 1\,eV. Despite the similarities in their structure, bonding, and composition, the four heterostructures host markedly different excitons, arising from distinct ELAs at the interfaces. For the PDI-WS$_2$ bilayer, we identify the formation of charge-transfer ILEs and hybrid excitons with binding energies near 600\,meV, small exciton Bohr radii (1.5-3\,nm), and long radiative lifetimes (0.1-5\,ns). These characteristics $-$ long lifetimes and tunable interactions $-$ make these hybrid systems advantageous for exploring many-body quantum phenomena in 2D materials, including correlated states~\cite{chowdhury2025}, excitonic insulators, and strongly-coupled light-matter systems.

\section{Results}

\begin{figure}[h!]
	\includegraphics[width=0.99\textwidth]{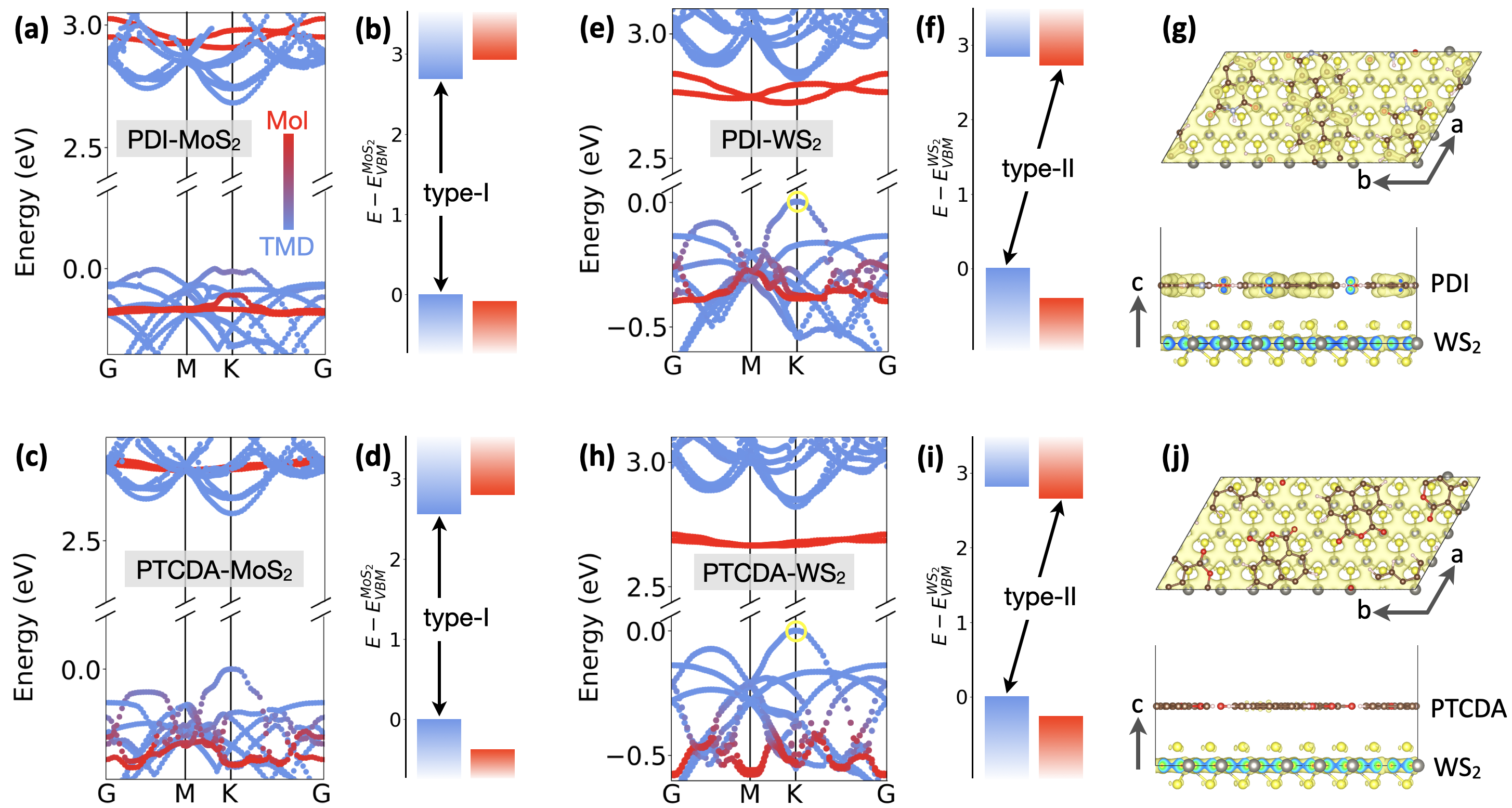}
	\centering
	\caption{Computed GW band structure for (a) PDI-MoS$_2$, (c) PTCDA-MoS$_2$, (e) PDI-WS$_2$, and (h) PTCDA-WS$_2$. Band colors result from the projection of interfacial orbitals onto  orbitals of the individual layers. Schematic of band edges for (b) PDI-MoS$_2$, (d) PTCDA-MoS$_2$, (f) PDI-WS$_2$, and (i) PTCDA-WS$_2$. DFT-calculated real-space squared wave functions of the VBM-HOMO at K point (shown by yellow circle) for (g) PDI-WS$_2$ and (j) PTCDA-WS$_2$ $-$ top and side views. Spin-orbit coupling effects are not included.}
	\label{fig2}
\end{figure}

Figure~\ref{fig2} shows our computed GW band structures for all four heterostructure bilayers, along with a schematic illustration of the ELA in each case. The colors denote the character of the quasiparticle bands, determined through the projection of Kohn-Sham eigenstates onto states of the individual layers: light blue bands indicate states localized on the TMD, while red bands correspond to states localized on the molecular adsorbates. Key energy gaps are summarized in Table~\ref{tab1}. We highlight that the GW gaps for the freestanding TMD monolayers are in good agreement with experimental and computed values reported previously~\cite{ramasubramaniam2012,qiu2013,qiu2016}.

\begin{table}[h!]
    \centering
    \begin{tabular}{l | c | c c c}
    \hline
     &            & \multicolumn{3}{c}{For Interfaces} \\ \cline{3-5}
     & GW (DFT) gap & VBM-CBM gap & HOMO-LUMO gap & Type \\
     \hline
     PDI monolayer & $4.06$ ($1.41$) &  &  & \\
     PTCDA monolayer & $4.21$ ($1.46$) &  &  & \\
     MoS$_2$ monolayer & $2.70$ ($1.75$) &  &  & \\
     WS$_2$ monolayer & $2.85$ ($1.89$) &  &  & \\
     \hline
     PDI-MoS$_2$ & $2.66$ ($1.38$) & $2.66$ ($1.75$) & $3.01$ ($1.38$) & I (I) \\
     PTCDA-MoS$_2$ & $2.64$ ($1.33$) & $2.64$ ($1.74$) & $2.93$ ($1.42$) & I (II) \\
     PDI-WS$_2$ & $2.72$ ($1.23$) & $2.82$ ($1.88$) & $2.92$ ($1.37$) & II (II) \\
     PTCDA-WS$_2$ & $2.67$ ($1.12$) & $2.82$ ($1.88$) & $2.88$ ($1.35$) & II (II) \\
     \hline
    \end{tabular}
    \caption{GW (DFT-PBE-D3) band gaps, in eV, computed for freestanding PDI, PTCDA, MoS$_2$, and the four bilayers. The gap between bands with a dominant molecular character is reported as ``HOMO-LUMO gap'' (HOMO: highest occupied molecular orbital; LUMO: lowest unoccupied molecular orbital), while the gap between bands with a dominant TMD character is reported as ``VBM-CBM gap'' (VBM: valence band maximum; CBM: conduction band minimum). All bilayer gaps and ELA values are reported for the K/K' point. Spin-orbit coupling effects are not included. The type of interface present in the heterostructure is mentioned in the last column.}
    \label{tab1}
\end{table}

Importantly, the electronic band structures of the hybrid bilayers cannot be simply predicted from a superposition of those of its constituents. They exhibit features that differ significantly from the nearly flat bands of isolated molecular monolayers and the characteristic dispersive bands of TMD monolayers (see Fig.~S1).

From our calculations, we make the following observations. First, in all four systems, energies of bands with dominant HOMO and LUMO character (shown in red) undergo significant renormalization in the bilayer, primarily due to non-local screening from the TMD. For instance, the PTCDA monolayer experiences a band gap renormalization of up to 1.28\,eV (1.33\,eV) when interfaced with MoS$_2$ (WS$_2$). This leads to a QP gap notably smaller than its isolated molecular monolayer value of 4.21\,eV and gas phase value of 4.70\,eV~\cite{sharifzadeh2012,refaely2013}. On the contrary, the TMD band gap remains largely unaffected, suggesting negligible renormalization from the weak dielectric screening of the molecular monolayer. As expected, our DFT-PBE calculations clearly underestimate the relevant gaps, do not capture the gap renormalization in the bilayer~\cite{neaton2006,refaely2013}, and, in some cases, fail to predict the correct type of interfacial ELA (see Table~\ref{tab1}).

Second, for the PDI-MoS$_2$, PTCDA-MoS$_2$, and PDI-WS$_2$ bilayers, the bands with dominant VBM character show reduced dispersion compared to those of freestanding TMD monolayers. This suggests hybridization between the TMD and molecular monolayer states, and is further supported by the presence of additional shoulder features (and purple intensity) in Figs.~\ref{fig2}a, \ref{fig2}c, and \ref{fig2}e. This brings us to the key distinction between PDI-WS$_2$ and PTCDA-WS$_2$: the real-space electron density associated with the VBM at the K point in PDI-WS$_2$ extends more significantly in the vertical direction across both layers (Fig.~\ref{fig2}g), while in PTCDA-WS$_2$, it remains more confined within the TMD layer (Fig.~\ref{fig2}j), closely resembling the behavior of the freestanding monolayer.

Notably, in the PDI-WS$_2$ bilayer, the bands with LUMO and LUMO+1 character exhibit substantial bandwidth broadening $-$ up to 90\,meV $-$ and the degeneracy of the LUMO band is lifted at $\Gamma$ and K. The broadening and degeneracy lifting reflect strong in-plane intermolecular interactions and enhanced hybridization. The effect is more pronounced in PDI than in PTCDA, consistent with the differing 2D molecular arrangements: the brick-wall packing of PDI favors closer proximity of functional groups and thus stronger interactions, whereas PTCDA adopts a more weakly-coupled herringbone geometry (see Fig.~S3).

Third, we emphasize the major role of the TMD monolayer in governing the ELA (Figs.~\ref{fig2}b, \ref{fig2}d, \ref{fig2}f, and \ref{fig2}i). In particular, MoS$_2$-based bilayers exhibit a type-I alignment, while WS$_2$-based bilayers display a type-II alignment. The computed DFT-PBE work function of MoS$_2$ is about 320\,meV larger than that of the WS$_2$ monolayer, which results in a substantially smaller LUMO-CBM offset in PDI-MoS$_2$ (305\,meV) compared to PDI-WS$_2$ bilayer (645\,meV). Although GW self-energy corrections to the LUMO band are similar in magnitude across all four systems, the outcome differs: in MoS$_2$-based systems, the LUMO band is mixed into the conduction bands of MoS$_2$, while in WS$_2$-based systems, it lies below the CBM of WS$_2$, resulting in type-II alignment. These predictions align well with our recent electrochemical transport measurements~\cite{chowdhury2025}.

While type-II alignments have been observed in phtalocyanine-TMD systems~\cite{adeniran2021,ulman2021}, where the fundamental gap was computed to be between the TMD CBM and the HOMO level, PDI-WS$_2$ and PTCDA-WS$_2$ bilayers exhibit a type-II alignment with the gap being between the molecular LUMO band and TMD VBM. We note that this kind of alignment has yet to be reported for a hybrid bilayer and has implications for charge separation, photocurrent polarity, and the spatial distribution of excess carriers under gating or doping.\\

\begin{figure}[h!]
	\includegraphics[width=0.99\textwidth]{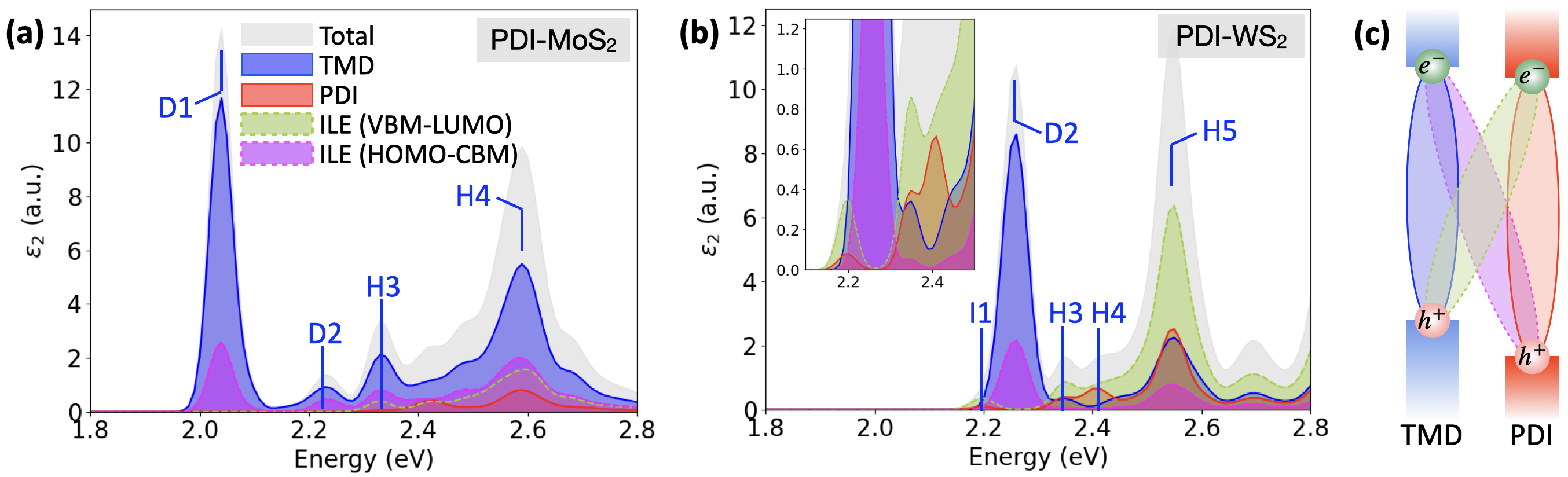}
	\centering
	\caption{For (a) PDI-MoS$_2$ and (b) PDI-WS$_2$ interfaces, the imaginary part of the dielectric function, $\epsilon_2$, is plotted as a function of the photon energy, showing the contributions from four distinct excitonic species, schematically represented in (c): TMD intralayer (blue), molecular intralayer (red), low-energy interlayer (green), and high-energy interlayer (magenta) excitons. The dominant nature of each peak is indicated on the figure with ``D'' for direct (intralayer), ``H'' for hybrid, and ``I'' for interlayer excitons.}
	\label{fig3}
\end{figure}

The unique electronic structure of these hybrid bilayer heterostructures also has implications for their photophysics. Our calculated GW-BSE optical absorption spectra for PDI-MoS$_2$ and PDI-WS$_2$ bilayers are shown in Figs.~\ref{fig3}a and \ref{fig3}b (gray curve), respectively, while corresponding spectra for PTCDA-MoS$_2$ and PTCDA-WS$_2$ heterostructures are presented in Fig.~S6. Using the scheme described in the SI~\cite{uddin2022}, the bilayer excited states are decomposed into different types of transitions $-$ those localized on the molecular (red) or TMD (blue) monolayer, and those with charge-transfer nature across the interface (green and magenta, respectively), as sketched in Fig.~\ref{fig3}c. This analysis enables us to attribute each spectral feature to a specific type of excitation and to understand the effects that modulate the excitonic properties of each monolayer within the bilayer systems. In the computed optical spectra in Fig.~\ref{fig3}, we identify and label multiple prominent features $-$ selected based on their high oscillator strength $-$ for discussion below. Additionally, for each state we compute the electron-hole correlation function, $\mathcal{F}_S$, representing the probability of finding the electron and the hole separated by a vector $\mathbf{r} = \mathbf{r}_e - \mathbf{r}_h$, and defined as
$ \mathcal{F}_S(\mathbf{r}) = \int_{\Omega} d^3\mathbf{r}_h | \Psi_S(\mathbf{r}_e = \mathbf{r}_h + \mathbf{r}, \mathbf{r}_h) |^2$, where $\Psi_S(\mathbf{r}_e,\mathbf{r}_h)$ is the normalized electron-hole wave function, $\mathbf{r}_e$ ($\mathbf{r}_h$) is the electron (hole) coordinate, and $\Omega$ the volume of the primitive cell~\cite{sharifzadeh2013,sharifzadeh2015}. In Table~\ref{tab2}, we report the excitation energy $\Omega^S$, exciton binding energy $E_b$, squared transition dipole moment $|\mu_S|^2$, exciton radius $a_x$, and zero-temperature radiative lifetime $\tau_0$ for selected bilayer excitons, comparing with those in the corresponding freestanding monolayers (see SI for details and definitions).

\begin{table}[h!]
    \centering
    \begin{tabular}{l | c c c c c c c}
         \hline
         System    & Index & $\Omega^S$ & $E_b$ & $|\mu_S|^2$  & $a_x$  & $\tau_0$ \\
                   &       & (eV)   & (meV)          & (a.u.)     & (\AA)   & (ps)     \\
         \hline
         PDI monolayer        & D1 & $2.61$ & $1450$ & $0.35 \times 10^5$ & $4.88$  & $0.05$ \\
         PTCDA monolayer     & D1 & $2.65$ & $1560$ & $0.16 \times 10^5$ & $4.96$  & $0.10$ \\
         MoS$_2$ monolayer   & D1 & $1.92$ & $780$  & $0.12 \times 10^5$ & $10.80$ & $0.18$ \\
         WS$_2$ monolayer    & D1 & $2.18$ & $670$  & $0.11 \times 10^5$ & $15.98$ & $0.18$ \\
         \hline
         PDI-MoS$_2$            & D1 & $2.03$ & $630$ & $0.70 \times 10^3$ & $13.83$ & $0.18$ \\ 
                       & D2 & $2.23$ & $430$ & $0.31 \times 10^2$ & $24.90$ & $3.68$ \\ 
                       & H3 & $2.32$ & $350$ & $0.13 \times 10^3$ & $33.11$ & $0.86$ \\ 
                       & H4 & $2.59$ & $110$  & $0.17 \times 10^2$ & $44.53$ & $5.74$ \\ 
         PDI-WS$_2$            & I1 & $2.16$ & $560$ & $0.24 \times 10^0$ & $16.23$ & $503.86$ \\ 
                       & D2 & $2.27$ & $550$ & $0.60 \times 10^3$ & $20.68$ & $0.19$ \\ 
                       & H3 & $2.34$ & $401$ & $0.95 \times 10^2$ & $28.98$ & $1.15$ \\ 
                       & H4 & $2.41$ & $381$ & $0.68 \times 10^2$ & $30.54$ & $1.57$ \\ 
                       & H5 & $2.54$ & $425$ & $0.14 \times 10^3$ & $30.07$ & $0.71$ \\ 
         \hline
    \end{tabular}
    \caption{For freestanding PDI, PTCDA, MoS$_2$, WS$_2$ monolayers, and PDI-MoS$_2$ and PDI-WS$_2$ interfaces, we report the BSE computed properties of the excited states, without spin-orbit coupling effects, including the type of transition (D - direct/intralayer; I - interlayer; H - hybrid), excitation energy ($\Omega^S$), the exciton binding energy ($E_b$), squared dipole moment ($|\mu_S|^2$), exciton radius ($a_x$), and radiative lifetime at zero-temperature ($\tau_0$). The indices refer to the peaks identified in Fig.~\ref{fig3}.}
    \label{tab2}
\end{table}

For the type-I PDI-MoS$_2$ bilayer, the lowest-energy computed excitation at 2.03\,eV corresponds to an intralayer MoS$_2$ exciton, where both the electron and hole are confined within the MoS$_2$ layer (Fig.~S7). This intralayer $1s$ exciton has a binding energy of 630\,meV, which is 150\,meV smaller than in the freestanding case, a renormalization similar in magnitude as that observed for monolayer MoS$_2$ on hBN~\cite{latini2017}. Excitonic features at higher energies correspond to both MoS$_2$ intralayer states (at 2.23\,eV) and hybrid excitons (at 2.32\,eV and above). Distinct from ILEs, hybrid excitons have mixed PDI-MoS$_2$ character and non-negligible charge-transfer nature, as evident from our computed optical spectra (Fig.~\ref{fig3}a) and the exciton wave function maps in Fig.~S7d. This pronounced delocalization in both lateral and vertical directions derives from the hybridized band structure (Fig.~\ref{fig2}a) and charge carrier delocalization, also observed in the PDI-WS$_2$ bilayer (Fig.~\ref{fig2}g).

\begin{figure}[h!]
	\includegraphics[width=0.85\textwidth]{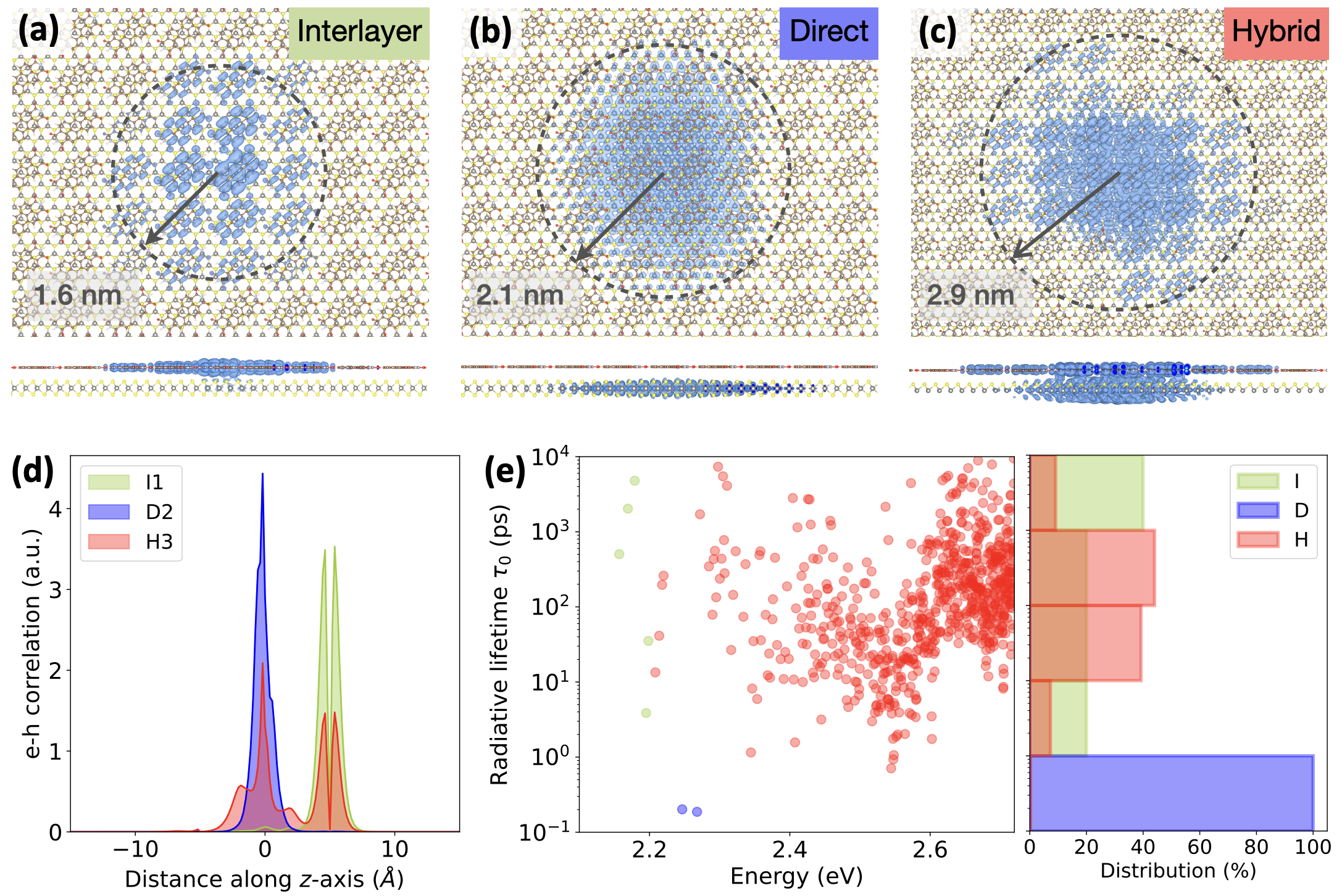}
	\centering
	\caption{For the PDI-WS$_2$ bilayer, we show the isosurface maps of the exciton wavefunctions (selected to include 98\% of the electron density), showing the electron probability density for the most probable hole positions: (a) I1 interlayer, (b) D2 WS$_2$ intralayer, and (c) H3 hybrid excitons. A $20 \times 10 \times 1$ supercell is used to generate these isosurface plots. The exciton Bohr radii (quantitatively defined in SI) are shown with the gray arrows. (d) Electron-hole correlation function along the $z$-axis. (e) Distribution of exciton radiative lifetimes with respect to the photon energy, with the intralayer, interlayer, and hybrid excited states shown in blue, green, and red, respectively.}
	\label{fig4}
\end{figure}

For the type-II PDI-WS$_2$ bilayer, Figure~\ref{fig4}d shows the electron-hole correlation function along the out-of-plane ($z$) direction for three distinct exciton types $-$ labeled I1, D2, and H3 $-$ centered around the WS$_2$ layer ($z=0$\,\AA), with the PDI layer at approximately $z=5$\,\AA. Figures~\ref{fig4}a-\ref{fig4}c display the lateral and vertical isosurface maps of electron probability densities for those three excitons, with the hole fixed at high-probability positions (see SI). The lowest-energy excitation at 2.16\,eV is an interlayer exciton, with electron localized in the PDI layer for a hole position in the WS$_2$ layer. In contrast, the high-intensity peak at 2.27\,eV corresponds to an intralayer WS$_2$ exciton with both the electron and hole confined in the TMD layer.

Excitons at 2.34\,eV and beyond correspond to hybrid excitons with mixed molecular and TMD character. Hybrid states H3 and H4 in Fig.~\ref{fig3}b comprise intralayer WS$_2$ excitations (blue) resonant with intermolecular PDI charge-transfer excited states (red). While the lowest-energy exciton in the isolated PDI monolayer at 2.61\,eV is strongly localized and Frenkel-like (Fig.~S5b), higher-energy exciton states are significantly more delocalized, as also observed in systems like ZnPc~\cite{ulman2021} and pentacene~\cite{cudazzo2015,alvertis2023}. Due to the renormalization of the PDI monolayer band gap and a comparable reduction in exciton binding energy at the interface, intermolecular charge-transfer excitons appear at similar energies as in the freestanding PDI layer. This results in a broader spatial extent of the exciton wave function across the molecular layer, relative to typical Frenkel excitons observed in molecular systems.\\

Our GW-BSE calculations for the four organic-TMD bilayers reveal marked tunability of their optoelectronic properties, stemming directly from the choice of molecular or TMD monolayer. Notably, substituting monolayer MoS$_2$ with WS$_2$ transforms the system from type-I to type-II, with significant implications. In PDI-MoS$_2$, the lowest-energy exciton is a tightly-bound intralayer MoS$_2$ exciton, whereas in PDI-WS$_2$, the lowest-energy excitation is an interlayer exciton, with the electron localized in the PDI layer and the hole in the WS$_2$. Interestingly, higher-energy hybrid excitons emerge in all four heterostructures (see Fig.~\ref{fig4}), not just for type-II heterostructures, but also in type-I, as recently reported in Ref.~\citenum{bennecke2024}. These mixed states exhibit partial charge-transfer character, which we can quantify via the electron-hole correlation function over the individual layers (see SI for details). This decomposition, yielding an effective charge-transfer percentage, ranges from 0\% (pure intralayer) to 100\% (pure interlayer). For PDI-WS$_2$, the lowest-energy interlayer exciton exhibits 96\% charge transfer, while hybrid states can carry up to 42\%, highlighting the diverse excitonic landscape accessible in such vdW heterostructures (Figs.~\ref{fig4}a-\ref{fig4}c).

\begin{figure}[h!]
	\includegraphics[width=0.95\textwidth]{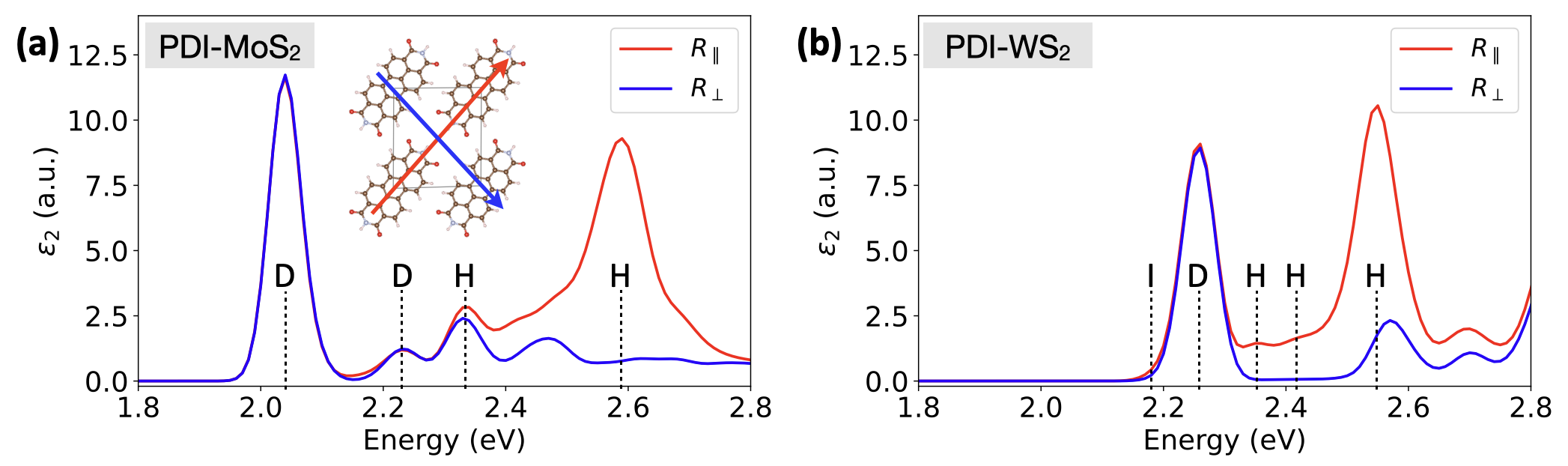}
	\centering
	\caption{For (a) PDI-MoS$_2$ and (b) PDI-WS$_2$ systems, we show the GW-BSE computed imaginary part of the dielectric function, $\epsilon_2$, as a function of the photon energy, for two orthogonal polarization directions (see inset). The prominent features are indicated by vertical dashed lines, with ``D'' for direct (intralayer), ``H'' for hybrid, and ``I'' for interlayer excitons.}
	\label{fig5}
\end{figure}

In addition, our GW-BSE absorption spectra, computed for two orthogonal polarization directions (Figs.~\ref{fig5}a and \ref{fig5}c), reveal that hybrid states in PDI-MoS$_2$ and PDI-WS$_2$ exhibit strong optical polarization anisotropy, in stark contrast with TMD intralayer excitons, whose oscillator strength is not altered much by the light polarization orientation~\cite{chowdhury2025}. For PDI-WS$_2$, we find an anisotropy ratio $\varphi$ of 92\%, 0.002\%, and 99\% for the I1, D2, and H3 excited states, respectively, defined as $\varphi = \frac{\mu_{\parallel}-\mu_{\perp}}{\mu_{\parallel}+\mu_{\perp}}$ where $\mu_{\parallel}$ and $\mu_{\perp}$ are the oscillator strengths for parallel and perpendicular polarizations, respectively.

The lifted degeneracy of the low-lying conduction bands of LUMO/LUMO+1 character in the PDI-WS$_2$ bilayer (Fig.~\ref{fig2}d) suggests that $n$-type doping may lead to electrons being localized on one of the two molecules in the PDI monolayer unit cell sublattice, rather than being a superposition across both. Our computed projected density of states (Fig.~S5) shows the lowest band of LUMO character is preferentially localized on one sublattice whereas the LUMO+1 band localizes on the other. This holds important consequences for doped systems, potentially influencing charge carrier mobility, charge transport pathways, and optical properties.

More broadly, our calculations for these four hybrid bilayers are corroborated by recent experimental observations~\cite{chowdhury2025}), where hybrid excitons in PDI-MoS$_2$ bilayers and interlayer and hybrid excitons in PDI-WS$_2$ bilayers were optically resolved. For PDI-WS$_2$, linearly polarized light parallel to the transition dipole moment of the molecules resulted in a strong optical signal, while perpendicular polarization yielded a weak optical response (Fig.~2e in Ref.~\citenum{chowdhury2025}). The strong polarization dependence of the optical spectrum of PDI-WS$_2$ and in particular of the interlayer ($\varphi_{\text{PL}} = \frac{I_{\text{PL}}^{\parallel}-I_{\text{PL}}^{\perp}}{I_{\text{PL}}^{\parallel}+I_{\text{PL}}^{\perp}} = 83$\%) and hybrid ($\varphi_{\text{PL}} = 97$\%) excitonic peaks $-$ largely absent in PTCDA-WS$_2$ $-$ underscore the sensitivity of optical response to molecular ordering, and confirms the role of the 2D brick-wall arrangement of PDI in tailoring excitonic behavior.\\

The interlayer excitons computed here in type-II organic-TMD bilayers have potential for Bose-Einstein condensation, similar to what has been reported before for other 2D materials and heterostructures~\cite{fogler2014,wang2019,gupta2020,liu2021}. A degenerate 2D Bose gas of excitons may form at temperatures below the degeneracy temperature $T_d$ when the mean inter-exciton distance $1/\sqrt{n}$, where $n$ is the number of excitons per area, and the exciton size (root-mean-square radius, $r_s$) satisfy $(1/\sqrt{n_c})^2 \approx \pi r_{s}^2$, where $n_c$ is the Mott critical density~\cite{palo2002,fogler2014}. The characteristic temperature $T_d$ at which excitons become degenerate can be estimated using the exciton effective mass $m_x$ and density $n_c$ via $k_B T_d \approx \frac{2\pi \hbar^2}{m_x}n_c$ with $k_B$, the Boltzmann constant~\cite{fogler2014}. Excitons with a small effective mass ($m_x$) and a large Mott critical density ($n_c$), which implies a small exciton radius, can increase $T_d$. In practice, excitons should have a long radiative lifetime ($\tau_0$) to maintain the condensate. In Table~\ref{tab3}, we report the exciton binding energy, Bohr radius, effective mass, Mott critical density, predicted T$_d$, and lifetime, for the ILE in PDI-WS$_2$ and PTCDA-WS$_2$ bilayers. These values are comparable to computed data for various exciton condensation candidates~\cite{fogler2014,nam1976,ulman2021}, as well as experimental results from systems where exciton condensation has been reported~\cite{stern2014,wang2019}.

\begin{table}[h!]
\centering
\begin{tabular}{l | c c c c c c}
\hline
Systems & $E_b$ (meV) & $a_x$ (nm) & $n_c$ (cm$^{-2}$) & $m_x$ ($m_0$) & $T_d$ (K) & $\tau_0$ \\
\hline
PDI-WS$_2$ (this work) & $560$ & $1.6$ & $1.73 \times 10^{13}$ & $4.61$ & $208$ & $504$\,ps \\
PTCDA-WS$_2$ (this work) & $590$ & $1.8$ & $1.37 \times 10^{13}$ & $11.50$ & $66$ & $77$\,ps \\
GaAs CQW~\cite{fogler2014,nam1976} & $4.2$ & $15$ & $10^{10}$ & $0.22$ & $3$ & $\sim \mu$s \\
MoS$_2$/hBN CQW~\cite{fogler2014} & $140$ & $1$ & $2 \times 10^{12}$ & $1.00$ & $100$ & $\sim \mu$s \\
ZnPc-MoS$_2$~\cite{ulman2021} & $690$ & $2.0$ & $2.17 \times 10^{13}$ & $11.57$ & $104$ & $860$\,ns \\
\hline
GaAs/AlGaAs CQW~\cite{stern2014} & $-$ & $25$ & $4\times 10^{10}$ & $-$ & $4.7$ & $60$\,ns \\
MoSe$_2$/hBN/WSe$_2$~\cite{wang2019} & $>100$ & $-$ & $0.74 \times 10^{12}$ & $-$ & $100$ & $1$\,ns \\
\hline
\end{tabular}
\caption{Computed properties of excitons in candidate systems for Bose-Einstein condensation, including the two systems of interest PDI-WS$_2$ and PTCDA-WS$_2$, GaAs CQW, MoS$_2$/hBN CQW (CQW: coupled quantum well), and ZnPc-MoS$_2$. These values are compared to experimental results from systems where exciton condensation has been reported, including GaAs/AlGaAs CQW and MoSe$_2$/hBN/WSe$_2$. We report the exciton binding energy ($E_b$), Bohr radius ($a_x$), Mott critical density ($n_c$), effective mass ($m_x$), degeneracy temperature ($T_d$), and radiative lifetime at 0\,K ($\tau_0$).}
\label{tab3}
\end{table}

In the PDI-WS$_2$ bilayer, we predict a momentum-direct ILE with high binding energy ($E_b = 560$\,meV), real-space localization ($a_x = 1.6$\,nm and $r_s = 1.36$\,nm; Fig.~\ref{fig4}a), and long radiative lifetime, ranging from 10\,ps to 5\,ns (Fig.~\ref{fig4}e). In comparison, the $1s$ exciton of WS$_2$ has a radius of 1.6\,nm in the freestanding monolayer, and extends up to 2.1\,nm ($r_s = 1.49$\,nm) in the bilayer (see Fig.~\ref{fig4}b). From Fig.~\ref{fig4}e, it is clear that radiative recombination of ILEs happens a few orders of magnitude slower than for intralayer excitons, reflecting the small oscillator strength (small overlap of wave functions) of spatially separated excitons. Hybrid excitons have intermediate radiative lifetimes, with a distribution centered around 0.1\,ns. 

We estimate the exciton effective mass, $m_x = m_e + m_h$~\cite{mattis1984,fogler2014}, where $m_e$ and $m_h$ are obtained as the second derivative of the CBM and VBM, respectively, with respect to wave vector $\mathbf{k}$. We find that in PDI-WS$_2$, $m_x = 4.61$\,$m_0$, dominated by $m_e = 4.19$\,$m_0$ ($m_0$ is the rest mass of an electron). In contrast, PTCDA-WS$_2$ exhibits an even larger mass, $m_x = 11.50$\,$m_0$, due to the flatter LUMO dispersion ($m_e = 11.09$\,$m_0$). These large effective masses and small radii imply high Mott densities and favorable $T_d$ (see Table~\ref{tab3}). The fact that our computed values are of the same order of magnitude as those reported for systems in which exciton condensates have been experimentally realized~\cite{stern2014,wang2019} strongly suggests that organic-TMD bilayers are not only highly tunable quantum materials but also may be promising platforms for exploring correlated phenomena.\\

We note that the longer lifetimes and dipolar character of ILEs in type-II interfaces such as PDI-WS$_2$ can also facilitate charge separation. The significant difference in effective masses between electrons and holes $-$ resulting from the flat LUMO in the molecule and the more dispersive VBM in the TMD $-$ favors anisotropic charge transport, with fast hole mobility and slow electron motion. When combined with the long-lived nature of ILEs, this separation can enable efficient charge dissociation and extraction following photoabsorption~\cite{lee2014}. Moreover, ILEs $-$ exhibiting spatial delocalization in real space $-$ will likely couple weakly to high-frequency phonons~\cite{alvertis2020}. Given that high-frequency phonons are responsible for the main heat loss in photovoltaics, minimizing the coupling of excitons to these modes can in turn greatly improve efficiencies~\cite{ghosh2024}. This further highlights the potential of organic-TMD van der Waals systems for optoelectronic applications with reduced heat losses.

\section{Conclusions}

In this study, we present a comprehensive and quantitative first-principles investigation of the electronic structure and excitonic properties of four 2D organic-TMD bilayers, using \textit{ab initio} GW-BSE calculations within the framework of many-body perturbation theory. We demonstrate the consequences of integrating 2D molecular monolayers into vdW heterostructures, and uncover how the interplay between molecular and TMD monolayers governs the emergent optoelectronic phenomena of these hybrid systems, not seen in conventional inorganic heterostructures.

Beyond the expected renormalization of the molecular monolayer gap at the interface, we demonstrate remarkable tunability in electronic energy level alignment and excitonic characteristics, governed by interfacial dielectric screening, hybridization, and charge transfer, that strongly vary with the TMD monolayer (MoS$_2$ vs. WS$_2$). In WS$_2$-based systems, we identify a momentum-direct ILE formed between the TMD VBM and the PDI/PTCDA LUMO bands. This ILE exhibits high binding energy ($E_b = 560$\,meV), compact real-space extent ($a_x = 1.6$\,nm), and long radiative lifetime ($\tau_0 = 504$\,ps), along with a remarkably high degree of charge transfer ($\sim 96$\%). These properties make such bilayers ideal candidates for exploring exciton transport and condensation, and offer key advantages for efficient charge separation in photovoltaic applications due to significantly different effective masses of electrons and holes, as well as reduced heat losses due to weak coupling of ILEs to high-frequency phonons. Furthermore, we identify hybrid excitons with pronounced anisotropic optical responses upon light polarization, observable in both type-I and type-II alignments, underscoring the richness of light-matter interaction in these systems.

Together, our findings reveal a highly tunable platform for studying excitonic physics in hybrid 2D materials and point to new strategies for  engineering exciton-based quantum and optoelectronic devices, based on molecularly engineered van der Waals heterostructures.


\begin{suppinfo}
The Supplementary Information is available free of charge.\\
\textit{Ab initio} calculations parameters; GW substrate screening approach; Embedding GW approach; Effect of strain on the electronic gap; LUMO/LUMO+1 degeneracy and projected density of states; Independent particle energy gap and exciton binding energies; Exciton decomposition; Two-particle correlation function; Exciton radii, lifetimes and effective masses; Analysis of optical spectra and excitonic properties (PDF).
\end{suppinfo}

\begin{acknowledgement}
The authors thank Su Ying Quek, and Steven G. Louie for stimulating discussions. The calculations in this work were primarily supported by the Center for Computational Study of Excited-state Phenomena in Energy Materials (C2SEPEM), funded by the US Department of Energy (DOE) under contract No.~DE-FG02-07ER46405. The Theory of Materials FWP at LBNL, funded by the DOE under contract No.~DE-AC02-05CH11231, supported the development of the theories and models. Computational resources are provided by the National Energy Research Scientific Computing Center (NERSC). Z.-F.L. acknowledges an NSF CAREER Award, DMR-2044552.
\end{acknowledgement}

\bibliography{referencias_sobraep.bib}

\end{document}